\documentclass[aps,prc,nofootinbib]{revtex4}
\usepackage{eurosym}
\usepackage{graphicx}
\usepackage{xcolor}

\begin{document}

\title{The charge and mass symmetry breaking in the $KK\bar{K}$ system}
\author{I. Filikhin$^1$, R. Ya. Kezerashvili$^{2,3,4}$, and B. Vlahovic$^1$}
\affiliation{\mbox{$^{1}$North Carolina Central University, Durham, NC, USA} \\
$^{2}$New York City College of Technology, The City University of New York,
Brooklyn, NY, USA\\
$^{3}$The Graduate School and University Center, The City University of New
York, New York, NY, USA\\
$^{4}$Long Island University, Brooklyn, NY, USA}

\begin{abstract}
In the framework of the Faddeev equations in configuration space, we
investigate the $K$(1460) meson as a resonant state of the $KK\bar{K}$
kaonic system. We perform calculations for the particle configurations $%
K^{0}K^{+}K^{-}$ and $K^{0}K^{+}\overline{{K}^{0}}$ within two models: the $%
ABC $ model, in which all three particles are distinguishable, and the $AAC$
model when two particles are identical. The models differ in their treatment
of the kaon mass difference and the attractive Coulomb force between the $%
K^{+}K^{-}$ pair. We found that the Coulomb shift adds over 1 MeV to the
three-body binding energy. The expected correction to the binding energy due
to mass redistribution from $AA$ to $AB$ is found to be negligible, up to a
maximum of 6\% of the relative mass correction. At the same time, the
symmetry of the wave function is distorted depending on the mass ratio
value. 
We found that the repulsive $KK$ interaction plays essential role in the binding energy of the $KK\bar K$ system and report the mass of 1461.8 or 1464.1 MeV for the neutral $K^{0}$(1460) and 1466.5 or 1468.8 MeV for the charged $K^{+}$(1460) resonances, respectively,
depending on the parameter sets for $KK$ and $K\bar{K}$ interactions.

\end{abstract}

\maketitle




\section{Introduction}

\label{sec:0} Few-body physics has received interest for decades. Since
1961, when the Faddeev equations \cite{Fad} in the momentum representation
were formulated and a few years later the Faddeev-Noyes equations in
configuration space were suggested \cite{Noyes1968}, special attention has
been given to three-body systems constituted by nucleons, mesons, two
nucleons and a meson, two mesons and a nucleon, quarks, and three-particle
cluster systems. At low energies, the general approach for solving the
three-body problem is based on the use of methods for studying the dynamics
of three particles in discrete and continuum spectra. Among the most
powerful approaches are the method of Faddeev equations in momentum \cite%
{Fad,Fad1} or coordinate \cite{Noyes1968,Noyes1969,Gignoux1974,FM} spaces.
However, the method of hyperspherical harmonics, the variational method in
the harmonic-oscillator basis, and the variational method complemented with
the use of explicitly correlated Gaussian basis functions has been
successfully employed for the solution of a few-body problem in atomic,
nuclear, high energy physics, and even in condensed matter physics \cite%
{RKez2019}.

Three-body systems can be composed of three identical particles ($AAA$ model), two
identical and the third one ($AAC$ model), and three non-identical particles ($ABC$ model). The $nnn$, $%
nns$, and $ssn$ baryons are examples of the quantum system with three and
two identical quarks, while 3$\alpha $ (cluster model for $^{12}$C), $ppn$($%
^{3}$He), and $nnp$($^{3}$H) nucleon systems are composed of three and two
identical particles \cite{FSV19}, respectively. One can also have systems
with a meson and two baryons or two mesons and a baryon, such as kaonic
clusters $NN{\bar{K}}$, ${\bar{K}\bar{K}}N$, $K{\bar{K}}N$, and $KK{\bar{K}}$%
, which are considered as systems with two identical or three
non-identical particle, depending on the configuration of the particles or
theoretical approach for the description of kaonic clusters. When two
identical particles are fermions or bosons the wave functions of the
three-body kaonic clusters are antisymmetric or symmetric, correspondingly,
with respect to the two identical particles exchange.

The kaonic clusters $NN{\bar{K}}$, ${\bar{K}\bar{K}}N$,\ and $K{\bar{K}}N$
were intensively studied in the framework of the Faddeev equations in
momentum and coordinate representation \cite%
{Ikeda2007A,Shevchenko2007A,Ikeda2007,Shevchenko2007B,Ikeda2009,Ikeda2010,Torres2010,5,K2015,FV,FVK20}%
. The Faddeev equations were used to study four-body kaonic clusters (see
review \cite{RKezNNNK}). It is a very challenging task to solve the Faddeev
equations exactly and usually introduce some reasonable approximations of
the Faddeev equations, such as the use of separable potentials,
energy-independent kernels, on-shell two-body scattering amplitudes, the Faddeev-type Alt-Grassberger-Sandhas
equations' \cite{AGS}. In the framework of the fixed-center approximation for the Faddeev equations, dynamically generated
three-body resonances formed via meson-meson and meson-baryon interactions were studied (see \cite {13,12,922,Oset2020} and references herein).

The $K$(1460) pseudoscalar was a subject of interest since the
middle of the seventieths. In 1976, the first high statistics study of the $%
K^{\pm }p\rightarrow K^{\pm }\pi ^{+}\pi ^{-}p$ process was carried out at
SLAC, using a 13 GeV incident $K^{\pm }$ beam \cite{Brandenburg}. The $%
J^{\pi}=0^{-}$ partial-wave analysis of the $\pi \pi K$ system in this
reaction led to the report of the first evidence for a strangeness-one
pseudoscalar meson with a mass of $\sim $ 1400 MeV and a width of $\sim $
250 MeV. A few years later the ACCORD collaboration \cite{DAUM1981} using
SLAC $K^{-}$ beam at 63 GeV investigated the diffractive process $%
K^{-}p\rightarrow K^{-}\pi ^{+}\pi ^{-}p$ and the data sufficient for
partial-wave analysis extending up to a mass of 2.1 GeV were collected. The
data analysis confirmed the existence of a broad $0^{-}$ resonance with a
mass of $\sim $ 1460 MeV. However, even the 2018 PDG \cite{PDG2018} did not list
the $K$(1460) as an "established particle". In the most recent studies of
the resonance structure in\ $D^{0}\rightarrow K^{\mp }\pi ^{\pm }\pi ^{\pm
}\pi ^{\mp }$ decays using $pp$ collision, data collected at 7 and 8 TeV
with the LHCb experiment \cite{LHCb}, and the precise measurements of the
$a_{1}^{+}$(1260), $K_{1}^{-}$%
(1270) and $K$(1460) resonances are made. Within a model-independent
partial-wave analysis performed for the $K$(1460) resonance, it is found that the
mass is roughly consistent with previous studies \cite{Brandenburg,
DAUM1981}. They showed the evidence for the K(1460) in ${\bar{K}}^{\ast
}(892)\pi ^{-}$ and $[\pi ^{+}\pi ^{-}]^{L=0}K^{-}$ channels and confirmed the resonant
nature of the $K$(1460) with the mass 1482.40$\pm$3.58$\pm$15.22 MeV and width 335.60$\pm$6.20$\pm$8.65 MeV. Thus, the $K$(1460) is listed in the 2020
PDG \cite{PDG2020} and in the most recent PDG \cite{PDG2022}.

Several theoretical models \cite%
{Izgur85,Longacre90,Albaladejo,Torres2011,RKSh.Ts2014,KezerasVail2015,Parganlija2017,SHY19,KezerasPRD2020,Torrez2021,Meisner2022}
considered the dynamic generation of the pseudoscalar $K$(1460) resonance as
three $K$ mesons: $KK{\bar{K}}$. The first time the resonance $K$(1460) was
considered as a $2^{1}S_{0}$ excitation of the kaon in a unified quark model
in Ref. \cite{Izgur85} and the mass 1450 MeV for this resonance was
obtained. In Ref. \cite{Longacre90} within the isobar assumption shown that the final-state interaction
leads to the formation of an exotic ($I(J^{\pi })=\frac{3}{2}(0^{-})$) $K^{+}K^{+}\overline{K^{0}}$ threshold enhancement with a mass around 1500 MeV and a width of approximately
200 MeV. Assuming isospin symmetry in the effective kaon-kaon interactions
that are attractive for the $K{\bar{K}}$ pair and repulsive for the $KK$
pair, 
in Ref. \cite{Torres2011} the $K$(1460) pseudoscalar resonance was studied
as the $KK{\bar{K}}$ system. The authors have performed the investigation of
the $KK{\bar{K}}$ system using two methods: the single-channel variational
approach in the framework of the model 
\cite{KanadaJido, JidoKanada}, and the Faddeev equations in momentum
representation. In the latter case, the authors determined the two-body on-shell $t$ matrices for $%
KK$ and $K{\bar{K}}$ interactions using the Bethe-Salpeter
equation in a couple-channel approach and considered the on-shell factorization method. Dynamic generation of the pseudoscalar $K$(1460) resonance
was considered in Ref. \cite{Albaladejo}, by studying interactions between
the $f_{0}(980)$ and $a_{0}(980)$ scalar resonances and the lightest
pseudoscalar mesons. In \cite{Torres2011} have included $\pi\pi K$ and $\pi\eta K$ channels. The same channels have been considered in an effective way in Ref. \cite{Albaladejo}.
In Ref. \cite{RKSh.Ts2014} using the single-channel
description of the $KK{\bar{K}}$ system in the framework of the
hyperspherical harmonics method was obtained a reasonable agreement with
experimental data for the mass of the $K$(1460) resonance. Authors \cite{Parganlija2017}, based on the extended version of the linear sigma model \cite{GellMann60,Gasiorowicz69,Ko94}, discussed to what extent it is possible to interpret the pseudoscalar meson $K(1460)$ ($I(J^{\pi })=\frac{1}{2}(0^{-})$) as excited $\bar{q}q$ states. In
a framework of the Faddeev equations in configuration space using the $AAC$ model $KK{\bar{K}}$
system was investigated in Ref. \cite{KezerasPRD2020}.
In Ref. \cite{SHY19}
the $KK{\bar{K}}$ system was considered based on the coupled-channel
complex-scaling method by introducing three channels $KK{\bar{K}}$, $\pi
\pi K$, and $\pi\eta K$. The resonance energy and width were determined
using two-body
potentials that fit two-body scattering properties. The model potentials
having the form of one-range Gaussians were proposed based on the
experimental information related to $a_{0}$ and $f_{0}$ resonances. In this
model, the $K\bar{K}$ interaction depends on the pair isospin. In
particular, the isospin triplet $K{\bar{K}}(I=1)$ interaction is essentially
weaker than the isospin singlet $K{\bar{K}}-K{\bar{K}}$ interaction in the
channel $\pi K-K{\bar{K}}(I=0)$.  The most recently in Ref. \cite{Meisner2022} the $KK{\bar{K}}$ three-body system in $\frac{1}{2}(0^{-})$ channel is studied in a formalism that satisfies two-body and three-body
unitarity using the isobar approach, where the two-body $K{\bar{K}}$ interaction is parametrised via the $f_{0}(980)$ and $a_{0}(980)$ poles.

There are the following physical particle configurations of the $KK{\bar{K}}$
system: $K^{0}K^{0}{\bar{K}}$, $K^{0}K^{+}K^{-}$, $K^{0}K^{+}\overline{{K}%
^{0}}$, $K^{+}K^{+}{\bar{K}}$. In these configurations, there are particles and
antiparticles, that have different masses, isospins, and electric charges.
The $K^{0}K^{0}\overline{{K}^{0}}$ and $K^{+}K^{+}{\bar{K}}$ configurations can
be considered in the framework of Faddeev equations as an $AAC$ model with
two identical particles. The configurations $K^{0}K^{+}K^{-}$ and $K^{0}K^{+}%
\overline{{K}^{0}}$ present the system with three non-identical particles
due to the different masses and isospins of mesons. The latter two
configurations must be considered within the Faddeev formalism as the $ABC$ system. Theoretical approaches for these systems utilizing average masses
of particles can lead to the loss of important information about the nature
of the systems. In particular, the formalization of the kaons as identical
particles allows us to discuss an exchange effect as was discussed for the $%
NN{\bar{K}}$ system \cite{YA07}.

We focus on the particle configurations $K^{0}K^{+}K^{-}$ and $K^{0}K^{+}%
\overline{{K}^{0}}$ for the bound three-body $KK{\bar{K}}$ system.
Previously, the kaonic $KK{\bar{K}}$ resonance is considered within the $AAC$
model using average kaon masses and disregarding the Coulomb force.
Considering the Coulomb potential or difference of the masses of kaons, we
propose the $ABC$ model for the description of the $K^{0}K^{+}K^{-}$ and $%
K^{0}K^{+}\overline{{K}^{0}}$ particle configurations rigorously within the
Faddeev equation in configuration space. One of the objectives
of this paper is to compare the binding energies of the $KK{\bar{K}}$ system
obtained within the $AAC$ and $ABC$ models using the same interaction
potentials. This allows us to draw a conclusion on the $AAC$ model's
validity. We show that the proper averaging procedure within the $ABC$
model leads to the $AAC$ model with a negligible correction of the binding
energy - up to a maximum of 6\% of the relative mass correction. For the
higher relative mass correction, the $AAC$ model is unacceptable, and one
must use the $ABC$ model.

In this paper we focus on the single-channel description of the $KK{\bar{K}}$ system in the absence
of the $\pi\pi K$ and $\pi\eta K$ inelastic channels. It
is worth noticing that despite its simplicity, the single-channel model can
reproduce the mass of the $K(1460)$ resonance. The smallness of the kinetic energy of the
kaons in a bound system in comparison with the kaon mass, makes possible the study of the single channel $KK{\bar{K}}$ using
a non-relativistic potential model. While the single-channel and non-relativistic potential model description clearly do not fully
represent reality, it still allows us to address the issues related to the charge and mass symmetry breaking in the $KK\bar{K}$ system and the validity of the $AAC$ model. In the $AAC$ and $ABC$ models, we are
using $s$-wave $K{\bar{K}}$ and $KK$ two-body potentials \cite{Torres2011}
and in the $ABC$ model experimental kaon masses, as the inputs.

This paper is organized as follows. In Secs. \ref{sec:2} and \ref{sec:2a} we
present the Faddeev equations in configuration space for $ABC$ and $AAC$
models and their application for the $K^{0}K^{+}K^{-}$ and ${K}^{0}K^{+}%
\overline{K^{0}}$ configurations. Averaged mass and potential approaches
which lead to the reduction of the $ABC$ model to the $AAC$ one are given in
Secs. \ref{sec:3} and \ref{sec:4}, respectively. Results of numerical
calculations for $AAC$ and $ABC$ models are presented in Sec. \ref{sec:5}.
The concluding remarks follow in Sec. \ref{sec:6}.

\section{Faddeev equations in configuration space}

\label{sec:2} The three-body problem can be solved in the framework of the
Schr\"{o}dinger equation or using the Faddeev approach in the momentum \cite%
{Fad,Fad1} or configuration \cite{Noyes1968,FM,K86} spaces. The Faddeev
equations in the configuration space have different forms depending on the
type of particles and can be written for: i. three non-identical particles; ii. three particles when two are identical; iii. three identical particles. The identical
particles have the same masses and quantum numbers. In the particle
configurations $K^{0}K^{+}K^{-}$ and $K^{0}K^{+}\overline{K^{0}}$ the $K^{+}$
and $K^{-}$, and $K^{0}$ and $\overline{K^{0}}$ have equal masses,
respectively. However, these particle configurations cannot be considered in
the framework of the $AAC\ $model because $K^{-}$ and $\overline{K^{0}}$ are
antiparticles of $K^{+}$ and $K^{0}$, respectively, hence, are non-identical
particles. Moreover, in the $K^{0}K^{+}K^{-}$ particle configuration the $%
K^{+}$ and $K^{-}$ are a particle and antiparticle, respectively, which have
the same masses but different charges. The non-identical particles are
un-exchangeable bosons. Thus, the $K^{0}K^{+}K^{-}$ and $K^{0}K^{+}\overline{%
K^{0}}$ particle configurations each consist of three non-identical
particles and must be treated within the \textit{ABC} model.

\subsection{Faddeev equations for the \textit{ABC} model}

In the Faddeev method in configuration space, alternatively, to the finding
the wave function of the three-body system using the Schr\"{o}dinger
equation, the total wave function is decomposed into three components \cite%
{Noyes1968,FM,K86}:
\begin{equation}
\Psi (\mathbf{x}_{1},\mathbf{y}_{1})=U(\mathbf{x}_{1},\mathbf{y}_{1})+W(%
\mathbf{x}_{2},\mathbf{y}_{2})+Y(\mathbf{x}_{3},\mathbf{y}_{3}).  \label{P}
\end{equation}%
Each Faddeev component corresponds to a separation of particles into
configurations $(kl)+i$, $i\neq k\neq l=1,2,3$. The Faddeev components are
related to their own set of the Jacobi coordinates $\mathbf{x}_{i}$ and $%
\mathbf{y}_{i}$, $i=1,2,3$. There are three sets of Jacobi coordinates. The
total wave function is presented by the coordinates of one of the sets shown
in Eq. (\ref{P}) for $i=1$. The mass-scaled Jacobi coordinates $\mathbf{x}%
_{i}$ and $\mathbf{y}_{i}$ are expressed via particle coordinates $%
\mathbf{r}_{i}$ and masses $m_{i}$ in the following form:
\begin{equation}
\mathbf{x}_{i}=\sqrt{\frac{2m_{k}m_{l}}{m_{k}+m_{l}}}(\mathbf{r}_{k}-\mathbf{%
r}_{l}),\qquad \mathbf{y}_{i}=\sqrt{\frac{2m_{i}(m_{k}+m_{l})}{%
m_{i}+m_{k}+m_{l}}}(\mathbf{r}_{i}-\frac{m_{k}\mathbf{r}_{k}+m_{l}\mathbf{r}%
_{l})}{m_{k}+m_{l}}).  \label{Jc}
\end{equation}%
In Eq. (\ref{P}), each component depends on the corresponding coordinate set
which is expressed in terms of the chosen set of mass-scaled Jacobi
coordinates. The orthogonal transformation between three different sets of
the Jacobi coordinates has the form:
\begin{equation}
\left(
\begin{array}{c}
\label{tran}\mathbf{x}_{i} \\
\mathbf{y}_{i}%
\end{array}%
\right) =\left(
\begin{array}{cc}
C_{ik} & S_{ik} \\
-S_{ik} & C_{ik}%
\end{array}%
\right) \left(
\begin{array}{c}
\mathbf{x}_{k} \\
\mathbf{y}_{k}%
\end{array}%
\right) ,\ \ C_{ik}^{2}+S_{ik}^{2}=1,
\end{equation}%
where
\[
C_{ik}=-\sqrt{\frac{m_{i}m_{k}}{(M-m_{i})(M-m_{k})}},\quad S_{ik}=(-1)^{k-i}%
\mathrm{sign}(k-i)\sqrt{1-C_{ik}^{2}}.
\]%
Here, $M$ is the total mass of the system. The components $U(\mathbf{x_{1}},%
\mathbf{y_{1}})$, $W(\mathbf{x}_{2},\mathbf{y}_{2})$, and $Y(\mathbf{\ x}%
_{3},\mathbf{y}_{3})$ satisfy the Faddeev equations in the coordinate
representation written in the form \cite{FM}:
\begin{equation}
\begin{array}{l}
(H_{0}+V_{23}(\mathbf{x_{1}})-E)U(\mathbf{x},\mathbf{y})=-V_{23}(\mathbf{%
x_{1}})(W(\mathbf{x}_{2},\mathbf{y}_{2})+Y(\mathbf{\ x}_{3},\mathbf{y}_{3}),
\\
(H_{0}+V_{13}(\mathbf{x_{2}})-E)W(\mathbf{x},\mathbf{y})=-V_{13}(\mathbf{%
x_{2}})(U(\mathbf{x}_{1},\mathbf{y}_{1})+Y(\mathbf{\ x}_{3},\mathbf{y}_{3}),
\\
(H_{0}+v_{12}(\mathbf{x_{3}})-E)Y(\mathbf{x},\mathbf{y})=-V_{12}(\mathbf{%
x_{3}})(U(\mathbf{x}_{1},\mathbf{y}_{1})+W(\mathbf{\ x}_{2},\mathbf{y}_{2}).%
\end{array}
\label{e:1}
\end{equation}%
Here, $H_{0}=-(\Delta _{\mathbf{x}}+\Delta _{\mathbf{y}})$ is the kinetic
energy operator with $\hbar ^{2}=1$ and $V_{kl}$ is the interaction
potential between the pair of particles $(kl)$, $i\neq k\neq l$. In the
system of equations (\ref{e:1}), the independent variables are $\mathbf{x}$
and $\mathbf{y}$ and can be chosen to be $\mathbf{x_{i}}$ and $\mathbf{y_{i}}
$, where $i$ is $1$ or $2$, or $3$. After that, the remaining
coordinates are expressed by the chosen coordinates according to the Jacobi
coordinates transformation (\ref{tran}).

\subsection{Faddeev equations for the \textit{AAC} model}

The system of Eqs. (\ref{e:1}) can be reduced to a simpler form for a case
of two identical particles when a particle $B$ in the $ABC$ model is replaced
by a particle $A$. In this case, for the Bose particles, the total wave
function of the system is decomposed into the sum of the Faddeev components $%
U$ and $W$ corresponding to the $(AA)C$ and $(AC)A$ types of rearrangements:

\[
\Psi =U+W+PW,
\]%
where $P$ is the permutation operator for two identical particles.
Consequently, Eqs. (\ref{e:1}) can be rewritten as follows \cite{K86}:
\begin{equation}
\begin{array}{l}
{(H_{0}+V_{AA}-E)U=-V_{AA}(W+PW)}, \\
{(H_{0}+V_{AC}-E)W=-V_{AC}(U+PW)}.%
\end{array}
\label{GrindEQ__1_}
\end{equation}%
In Eqs. (\ref{GrindEQ__1_}) ${V_{AA}}$ and ${V_{AC}}$ are the interaction
potentials between identical and non-identical particles, respectively.


\section{ The $ABC$ model versus $AAC$ for $K^{0}K^{+}K^{-}$ and ${K}^{0}K^{+}%
\overline{K^{0}}$ particle configurations}

\label{sec:2a} We study the $KK{\bar{K}}$ system using the available $s$%
-wave effective phenomenological $KK$ and $K{\bar{K}}$ potentials \cite%
{Torres2011} and considering the $K$ and ${\bar{K}}$ kaon's experimental
masses and charges\ based on Ref. \cite{PDG2022}. This leads to the
consideration of the $K(1460)$ resonance according to the following neutral
or charged particle configurations: $K^{0}K^{0}{\bar{K}}$, $K^{0}K^{+}K^{-}$, $%
K^{+}K^{+}{\bar{K}}$, $K^{0}K^{+}\overline{K^{0}}$. For the description of the $KK{%
\bar{K}}$ system within the $ABC$ and $AAC$ models, we focus, as on the
representative, on the following two particle configurations: $K^{0}K^{+}K^{-}$
and $K^{0}K^{+}{\bar{K}}^{0}$. The configuration $K^{0}K^{+}K^{-}$ includes
the Coulomb interaction and $K^{+}$ and $K^{-}$ have the same masses but are
non-identical that make three particles distinguishable. The configuration $%
K^{0}K^{+}\overline{{K}^{0}}$ does not include the Coulomb interaction and
two particles $K^{0}$ and $\overline{K^{0}}$ have the same masses but are
non-identical. Therefore, the exact treatment of the $K^{0}K^{+}K^{-}$ and $%
K^{0}K^{+}\overline{K^{0}}$ particle configurations must be done within the $%
ABC$ model. We consider $s$-wave pair potentials. Hence, the bound state
problem for the $KK{\bar{K}}$ system should be formulated using the Faddeev
equations in the $s$-wave approach \cite{K2015}. In the $s$-wave approach
for the $K^{0}K^{+}K^{-}$ particle configuration Eqs. (\ref{e:1}) reads:
\begin{equation}
\begin{array}{l}
(H_{0}+v_{K^{0}K^{+}}+v_{C}^{\mathcal{U}}-E)\mathcal{U}=-v_{K^{0}K^{+}}(%
\mathcal{W}+\mathcal{Y}), \\
(H_{0}+v_{K^{+}K^{-}}+v_{C}^{\mathcal{W}}-E)\mathcal{W}=-v_{K^{+}K^{-}}(%
\mathcal{U}+\mathcal{Y}), \\
(H_{0}+v_{K^{0}K^{-}}+v_{C}^{\mathcal{Y}}-E)\mathcal{Y}=-v_{K^{0}K^{-}}(%
\mathcal{U}+\mathcal{W}).%
\end{array}
\label{eq:22}
\end{equation}%
In Eqs. (\ref{eq:22}) $v_{K^{0}K^{+}}$, $v_{K^{+}K^{-}}$ and $v_{K^{0}K^{-}}$
are the $s$-wave $KK$ and $K{\bar{K}}$ potentials, while $v_{C}^{U}$, $%
v_{C}^{\mathcal{W}}$ and $v_{C}^{\mathcal{Y}}$ are the components of the
Coulomb potential related to the $K^{+}K^{-}$ electrostatic interaction depending on
the mass-scaled Jacobi coordinate of the corresponding Faddeev components $%
\mathcal{U}$, $\mathcal{W}$, and $\mathcal{Y}$. A consideration of the
Coulomb interaction in the framework of the Faddeev formalism is a
challenging problem \cite{Mer80}. Following \cite{FSV19}, we consider the
Coulomb $K^{+}K^{-}$ interaction included on the left-hand side of the
Faddeev equations (\ref{eq:22}) as a perturbation.

\begin{figure}[h]
\centering
\includegraphics[width=17 pc]{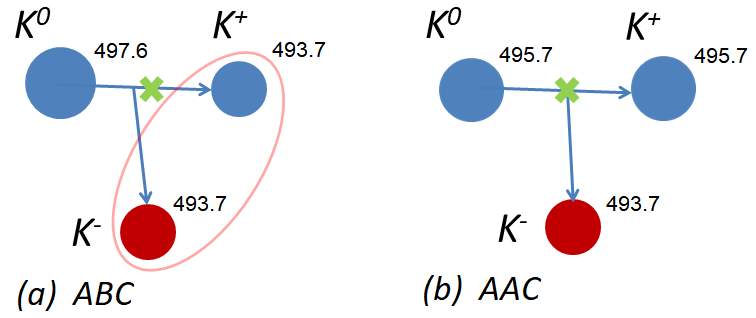}
\caption{ The schematic represents the reduction of the $ABC$ model ($a$) to $AAC$
model ($b$). The kaons and antikaon of the $K^{0}K^{+}K^{-}$ particle
configuration are shown in blue and red colors, respectively, together with
the experimental masses. The kaon pair with the Coulomb interaction is
encircled by the oval. The crosses indicate the position of the
middle point between $A$ and $B$ particles and two $A$ particles in the $ABC$
and $AAC$ models, respectively. The Jacobi coordinates related to
configurations $(AB)C$ and $(AA)C$ are shown by arrows. }
\label{fig2c}
\end{figure}

The Coulomb attraction in the $K^{0}K^{+}K^{-}$ violates the $AAC$ symmetry
and makes three kaons distinguishable. If one neglects the Coulomb
attraction in the $K^{0}K^{+}K^{-}$ in Eqs. (\ref{eq:22}), even the equality of $K^{+}$ and $K^{-}$ masses does not allow to treat the $K^{0}K^{+}K^{-}$
configuration in the framework of the $AAC$ model. However, if one considers
$K^{0}$ and $K^{+}$ as identical particles with masses equal to the average
of their masses and neglects the Coulomb attraction, $K^{0}K^{+}K^{-}$ can
be considered within the $AAC$ model. A schematic for the $K^{0}K^{+}K^{-}$
is presented in Fig. \ref{fig2c} when it is treated as $ABC$ and $AAC$
models. Within the $AAC$ model the Faddeev equations (\ref{GrindEQ__1_}) for
two identical particles in the three-body $K^{0}K^{+}K^{-}$ system for the $s
$-wave interparticle interactions can be written in the following form:
\begin{equation}
\begin{array}{l}
(H_{0}+v_{K^{0}K^{+}}-E)\mathcal{U}=-v_{K^{0}K^{+}}(1+P)\mathcal{W}\;, \\
(H_{0}+v_{K^{+}K^{-}}-E)\mathcal{W}=-v_{K^{+}K^{-}}(U+P\mathcal{W})\;.%
\end{array}
\label{eq:11}
\end{equation}%
In Eqs. (\ref{eq:11}) the $\mathcal{U}$ and $\mathcal{W}$ are the $s$-wave
Faddeev components of the wave function, and the exchange operator $P$ acts
on the particles' coordinates only.

The $K^{0}K^{+}\overline{K^{0}}$ particle configuration also can be
considered within the $AAC$\ model, if one considers $K^{0}$ and $K^{+}$ as
identical particles with masses equal to the average of their masses. In
this case, Eqs. (\ref{eq:11}) describe $K^{0}K^{+}\overline{K^{0}}$
configuration with only difference that $v_{K^{+}K^{-}}$ interaction should
be replaced by the $v_{K^{0}\overline{K^{0}}}$ potential.

%
%
%
%


\section{The mass difference of kaons: from the $AAC$ to $ABC$ model}

\label{sec:3} In the previous studies $KK{\bar{K}}$ system was considered
within the $AAC$ model. Such consideration is valid if we ignore the
difference between $K^{+}$ and $K^{0}$ masses and the Coulomb interaction
between charged kaons. As the first step for a realistic consideration, we
are using the experimental kaon masses instead of the average kaon mass used
in the $AAC$ model. After that, we consider that the $AB$ pair is the kaonic
pair and antikaon is considered as the particle $C$. Within this $ABC$
model, the masses of $A$ and $B$ kaons are varied around the average value
of the kaon mass ${m}=(m_{A}+m_{B})/2$. These variations have
different signs for the $A$ and $B$ kaons. In other words, we consider the $%
ABC$ model with variable masses of the $A$ and $B$ kaons but keeping the sum
of masses of the $AB$ kaon pair constant. This approach allows us to
understand how the binding energy of the $ABC$ is sensitive to the variation
of the $A$ and $B$ masses.

Consider the particles in the $ABC$ model, where $C$ is the antiparticle,
have masses $m_{A}$, $m_{C}$, and $m_{B}$, and can be numbered manually as
\begin{equation}
\begin{array}{l}
m_{1}=(m_{A}+m_{B})/2+m_{A}/2-m_{B}/2=m+\Delta m, \\
m_{2}=(m_{A}+m_{B})/2+m_{B}/2-m_{A}/2=m-\Delta m, \\
m_{3}=m_{C},%
\end{array}%
\label{m12}
\end{equation}%
where $\Delta m=(m_{A}-m_{B})/2$ is small and kaons are particles 1 and 2
and the antikaon is particle 3. Following Friar at al. \cite{F90}, we write the
kinetic energy operator $\hat{H_{0}}$ in terms of the individual momenta of the particles
in the center-of-mass as:
\begin{equation}
\hat{H_{0}}=\sum_{i=1,2,3}\frac{\pi _{i}^{2}}{2m_{i}}\approx \frac{\pi
_{3}^{2}}{2m_{3}}+\frac{\pi _{1}^{2}}{2m}+\frac{\pi _{2}^{2}}{2m}-\frac{%
\Delta m}{m}\frac{\pi _{1}^{2}}{2m}+\frac{\Delta m}{m}\frac{\pi _{2}^{2}}{2m}%
.  \label{cm}
\end{equation}%
This expression follows from the Taylor series for the $1/m$. In the first
order perturbation theory, the correction for the energy can be presented as
$\langle {\hat{H_{0}}}\rangle \approx \langle \hat{H_{0}}^{\Delta m=0}\rangle +\langle \Delta
H_{0}\rangle $, where $\langle \hat{H_{0}}^{\Delta m=0}\rangle=\langle \Psi ^{R}| \frac{\pi
_{3}^{2}}{2m_{3}}+\frac{\pi _{1}^{2}}{2m}+\frac{\pi _{2}^{2}}{2m}|\Psi ^{R}\rangle $ and 
\begin{equation}
\langle \Delta H_{0}\rangle =\langle \Psi ^{R}|\frac{\Delta m}{m}\frac{\pi
_{2}^{2}}{2m}-\frac{\Delta m}{m}\frac{\pi _{1}^{2}}{2m}|\Psi ^{R}\rangle .
\label{D}
\end{equation}%
Here, the $\Psi ^{R}$ is the coordinate part of the wave
function $\Psi =\eta _{isospin}\otimes \Psi ^{R}$.

Within the $AAC$ model, $\langle \Delta H_{0}\rangle =0$, due to $\Delta m=0$.
In the $ABC$ model, this relation
is approximately satisfied
\begin{equation}
\langle \Delta H_{0}\rangle \approx 0 .
\label{D1}
\end{equation}%
The possible
value for the $\Delta m$ is restricted so that $|\Delta m/m|\ll 1$. The
linear approximation Eq. (\ref{cm}) is applicable when we consider only the
first two terms of the Taylor series for the function $1/{m}$ near the point
$m=1$. The next quadratic terms of the expansion cannot be compensated similarly to Eq. (\ref{D1}) due to the alternating series of the Tailor expansion: $m/m_{i}=\sum_{n}(-1)^{n}(\Delta
m/m)^{n}$, $i=1,2$.

Therefore, we can assume that the symmetrical variation of kaons' masses described by Eq. (\ref{m12}) for the $ABC$ model
does not lead to a significant change in  the $AAC$ binding energy  when $|\Delta m/m|\ll 1$. This assumption follows from the compencation effect for the three-body Hamiltonian expressed by Eqs. (\ref{D})-(\ref{D1}).

\section{Coulomb interaction: from the $ABC$ to $AAC$ model}

\label{sec:4} Let us ignore the repulsive potential acting between the $%
K^{0} $ and $K^{+}$ kaons. Such truncation shifts the three-body energy to a
fixed value. After that the equation for the Faddeev component $\mathcal{U}$
in Eqs. (\ref{eq:22}) is eliminated. Also, we can neglect the Coulomb
interaction and consider only the nuclear interaction between kaons and
antikaon. The corresponding system of equations reads:
\begin{equation}
\begin{array}{l}
(H_{0}+v_{2}-E)\mathcal{W}=-v_{2}\mathcal{Y}, \\
(H_{0}+v_{3}-E)\mathcal{Y}=-v_{3}\mathcal{W},%
\end{array}
\label{eq:222}
\end{equation}%
where, for simplicity, we denoted $v_{K^{+}K^{-}}=v_{2}$ and $%
v_{K^{0}K^{-}}=v_{3}$. Assuming that the difference $\Delta
v=|(v_{3}-v_{2})/2|$ of the potentials $v_{2}$ and $v_{3}$ is small one can
introduce the average potential

\begin{equation}
\overline{v}=\frac{(v_{2}+v_{3})}{2},
\end{equation}%
so that

\begin{equation}
v_{2}=\frac{(v_{2}+v_{3})}{2}+\frac{(v_{2}-v_{3})}{2},\quad v_{3}=\frac{%
(v_{2}+v_{3})}{2}+\frac{(v_{3}-v_{2})}{2},
\end{equation}
and rewrite Eqs. (\ref{eq:222}) in the form

\begin{equation}
\begin{array}{l}
(H_{0}+\overline{v}+\Delta v-E)\mathcal{W}=-v_{2}\mathcal{Y}, \\
(H_{0}+\overline{v}-\Delta v-E)\mathcal{Y}=-v_{3}\mathcal{W}.%
\end{array}
\label{eq:333}
\end{equation}%
In the first order of the perturbation theory, by averaging Eqs. (\ref%
{eq:333}), one obtains
\begin{equation}
\begin{array}{l}
\langle \mathcal{W}_{0}|(H_{0}+\overline{v}+\Delta v-E)|\mathcal{W}%
_{0}\rangle =-\langle \mathcal{W}_{0}|(\overline{v}+\Delta v)|\mathcal{Y}%
_{0}\rangle , \\
\langle \mathcal{Y}_{0}|(H_{0}+\overline{v}-\Delta v-E)|\mathcal{Y}%
_{0}\rangle =-\langle \mathcal{Y}_{0}|(\overline{v}-\Delta v)|\mathcal{W}%
_{0}\rangle ,%
\end{array}
\label{eq:2222}
\end{equation}%
where $\mathcal{W}_{0}$ and $\mathcal{Y}_{0}$ are the solutions of Eqs. (\ref%
{eq:333}) with the potential $\overline{v}$, when $\Delta v$ is omitted,
that gives the energy of $E_{0}$. Therefore, from (\ref{eq:2222}) we obtain
\begin{equation}
\begin{array}{l}
\langle \mathcal{W}_{0}|E_{0}+\Delta v-E)|\mathcal{W}_{0}\rangle =-\langle
\mathcal{W}_{0}|\Delta v|\mathcal{Y}_{0}\rangle , \\
\langle \mathcal{Y}_{0}|(E_{0}-\Delta v-E)|\mathcal{Y}_{0}\rangle =-\langle
\mathcal{Y}_{0}|(-\Delta v)|\mathcal{W}_{0}\rangle .%
\end{array}
\label{eq:22a}
\end{equation}%
For equal kaon masses the functions $\mathcal{W}_{0}$ and $%
\mathcal{Y}_{0}$ are the same and, therefore, $\langle \mathcal{W}%
_{0}|\Delta v|\mathcal{W}_{0}\rangle =\langle \mathcal{Y}_{0}|\Delta v|%
\mathcal{Y}_{0}\rangle $ and $\langle \mathcal{W}_{0}|\Delta v|\mathcal{Y}%
_{0}\rangle =\langle \mathcal{Y}_{0}|\Delta v|\mathcal{W}_{0}\rangle $. By
adding the algebraic equations in (\ref{eq:22a}), we
obtain $E=E_{0}$. In other words, in the first order of perturbation theory,
the average potential gives $E=E_{0}$.

Now let's assume that masses of $K^{0}$ and $K^{+}$ are equal to their
average mass. The Coulomb potential is a perturbation with respect to the
strong kaon-kaon interaction. The Coulomb potential can be treated within the given above
scheme. One can denote the Coulomb part of the $K^{+}K^{-}$ potential as $%
v_{C}^{\mathcal{W}}$. The $v_{C}^{\mathcal{W}}$ is proportional to the $%
\frac{1}{x}$ and $v_{C}^{\mathcal{Y}}$ ($v_{C}^{\mathcal{U}}$) is
proportional to the $\frac{1}{x^{\prime \prime }}$ ($\frac{1}{x^{\prime }}$%
). Here, the mass-scaled Jacobi coordinate $x^{\prime }=|\mathbf{x_{1}}|$
corresponds to the $\mathcal{U}$ channel and is expressed by coordinates $x=|%
\mathbf{x_{2}}|$ and $y=|\mathbf{y_{2}}|$ of the channel $\mathcal{W}$. The $%
x^{\prime \prime }=|\mathbf{x_{3}}|$ is the coordinate in the channel $%
\mathcal{Y}$ conjugated to the $\mathcal{W}$ channel and is expressed via
coordinates $x$ and $y$ of the channel $\mathcal{W}$ (see Eq. (\ref{e:1})).

The average potential is $\overline{v}_{C}=(v_{C}^{\mathcal{W}}+v_{C}^{%
\mathcal{Y}})/2$. The $\Delta v_{C}=(v_{C}^{\mathcal{W}}-v_{C}^{\mathcal{Y}%
})/2$ defines the difference between the channels' potentials $v_{C}^{\mathcal{W}}$ and $v_{C}^{\mathcal{Y}}$. One can repeat the procedure
presented by Eqs. (\ref{eq:2222})-(\ref{eq:22a}) and finally for the $%
K^{0}K^{+}K^{-}$ with the Coulomb interaction obtains the following equation
that corresponds to the $AAC$\ model:
\begin{equation}
\begin{array}{l}
(H_{0}+v_{KK}+\overline{v}_{C}^{\mathcal{U}}-E)\mathcal{U}=-v_{KK}(1+P)%
\mathcal{W}, \\
(H_{0}+v_{KK^{-}}+\overline{v}_{C}^{\mathcal{W}}-E)\mathcal{W}%
=-v_{KK^{-}}(U+P\mathcal{W}),%
\end{array}
\label{eq:11a}
\end{equation}%
where
\begin{equation}
\begin{array}{l}
v_{KK}=v_{K^{0}K^{+}},\quad \overline{v}_{C}^{\mathcal{U}}=-n/x^{\prime },
\\
v_{KK^{-}}=v_{K^{0}K^{-}},\quad \overline{v}_{C}^{\mathcal{W}%
}=-n(1/x^{\prime \prime }+1/x)/2,%
\end{array}
\label{eq:11d}
\end{equation}%
and the masses of the kaons are equal to their average mass. The $n$ is the
Coulomb charge parameter.


\section{Numerical Results}

\label{sec:5}

\subsection{Interaction potentials}

In the present work, we use the $s$-wave effective potentials for ${K}{K}$
and $K{\bar{K}}$ interactions from Ref. \cite{Torres2011}. The $K{\bar{K}}$
is an attractive interaction that makes the $KK{\bar{K}}$ system bound, while the $%
KK $ interaction is described by a repulsive potential. The $K{\bar{K}}$ and $KK$
potentials are written in the form of one range Gaussian:
\begin{equation}
v_{{\bar{K}}K}(r)=v_{0}exp((-r/b)^{2}),
\end{equation}%
where $v_{0}$ and $b$ are the strength (depth) and the range of the potential, respectively. In
calculations are used two model potentials A and B with the set of
parameters listed in Table \ref{tab-1R}. 

\begin{table}[t]
\caption{The parameter sets A and B for $%
KK $ and $K\bar{K}$ interactions.}
\label{tab-1R}\centering
\par

\begin{tabular}{ccccc}
\hline\hline
\multicolumn{5}{c}{Parameters of potential} \\ \hline
& \multicolumn{2}{c}{Set A ($b=0.66$ fm)} & \multicolumn{2}{|c}{Set B ($%
b=0.47$ fm)} \\ \cline{2-5}
Interaction & $v_{0}$ ($I=0$), MeV & $v_{0}$ ($I=1$), MeV &
\multicolumn{1}{|c}{$v_{0}$ ($I=0$), MeV} & $v_{0}$ ($I=1$), MeV \\ \hline
$K\overline{K}$ & $-630.0-210i$ & $-630.0-210i$ & \multicolumn{1}{|c}{$%
-1155.0-283i$} & $-1155.0-283i$ \\
$KK$ & 0 & 104 & \multicolumn{1}{|c}{0} & 313 \\ \hline\hline
\end{tabular}

\end{table}

\subsection{From the $AAC$ to $ABC$ model: Effects of kaons mass difference and
Coulomb force}

Within the theoretical formalism presented in the previous sections, we
calculate the binding energies $E_{3}$ of the $K^{0}K^{+}K^{-}$ and ${K}%
^{0}K^{+}\overline{K^{0}}$ and  $E_{2}^{K{\bar{K}}}$ of the
bound $K{\bar{K}}$ pairs. The three-particle energy $E_{3}(V_{KK}=0)$, due to the kaon-antikaon interactions, but with the omitted interaction between two kaons, is another significant characteristic of the three-particle kaonic system. An analysis of the $E_{3}(V_{KK}=0)$ shows that the repulsive $KK$ interaction plays essential role in the resonance energy.

The results of calculations of these energies for the $%
K^{0}K^{+}K^{-}$ and ${K}^{0}K^{+}\overline{K^{0}}$ particle configurations
in the $AAC$ and $ABC$ models are presented in Table \ref%
{tab-3R}.

The analysis of the results leads to the following conclusions:

i. the binding energies $E_{3}$ and $E_{3}(V_{KK}=0)$ of $K^{0}K^{+}K^{-}$
and ${K}^{0}K^{+}\overline{K^{0}}$ calculated in the $AAC$ model, with the
average kaon masses $\overline{m}_{K}=495.7$ MeV, and 
$ABC$ model are
the same. Thus, the difference of the kaon masses does not affect $E_{3}$
and $E_{3}(V_{KK}=0)$: the mass distinguishability is not important for $%
E_{3}$ and $E_{3}(V_{KK}=0)$ energies when the Coulomb interaction is
neglected. The difference of 2.5 MeV for $E_{3}$ and 2.3 MeV for $%
E_{3}(V_{KK}=0)$ is related to the parameter sets A and B for $KK$ and $K{%
\bar{K}}$ interactions, respectively;

ii. the mass distinguishability has a small effect on the binding energy of $%
K{\bar{K}}$ pairs;

iii. the consideration of the Coulomb interaction in the framework of the $ABC$\
model leads to an increase of the binding energy $E_{3}$ and the energy $%
E_{2}^{K^{+}K^{-}}$ of the bound $K^{+}K^{-}$ pair in the $K^{0}K^{+}K^{-}$
particle configuration for the parameter sets A and B for $KK$ and $K{\bar{K}}$
interactions, respectively;

iv. the repulsive $KK$ interaction plays an essential role in the binding energy of the $KK\bar{K}$ system. The comparison of the $E_{3}$ and $E_{3}(V_{KK}=0)$ of $K^{0}K^{+}K^{-}$ and ${K}^{0}K^{+}\overline{K^{0}}$
energies shows that contribution of the repulsive $KK$ interaction decreases the three-particle binding energy by about $38\%$ and $25\%$ for the set of parameters A and B, respectively;

v. finally, the mass of neutral resonance $K(1460)$ in the $ABC$ ($%
K^{0}K^{+}K^{-}$) model is 1464.1 Mev and 1461.8 MeV for the parameter sets
A and B for $KK$ and $K{\bar{K}}$ interactions, respectively. The mass of
the charged resonance $K^{+}(1460)$ in the $ABC$ (${K}^{0}K^{+}\overline{%
K^{0}}$) model is 1468.8 MeV and 1466.5 MeV for the parameter sets A and B
for $KK$ and $K{\bar{K}}$ interactions, respectively. Our
results for the mass of neutral and charged $K$(1460) resonance obtained
within the $ABC$ model are in reasonable agreement with the reported
experimental value of the $K$(1460) mass 
\cite{PDG2022}.

Note, in contrast to the mass of the $K(1460)$ resonance, calculations of the
width using the Faddeev equations for the $AAC$ model in momentum
representation ($\Gamma =50$ MeV) \cite{Torres2011} and configuration
space ($\Gamma =104$ MeV) \cite{KezerasPRD2020}, variational method ($\Gamma =110$ MeV) \cite{Torres2011}, hyperspherical harmonics method ($\Gamma =49$ MeV) \cite{RKSh.Ts2014} did not reproduce the
quite sizeable experimental width, 335.60$\pm ${6.20}$\pm $8.65 MeV
\cite{LHCb,PDG2022}. In Ref. \cite{Longacre90} reported the width of approximately 200 MeV for the ${K}^{+}K^{+}\overline{K^{0}}$ resonance and \cite{Albaladejo} presented the estimation for the width of $\Gamma \geq 100$ MeV for $K(1460)$ resonance.
The study of the $KK\bar{K}$ within the non-perturbative three–body dynamics did not calculate the width for this system.

We calculate the width follow \cite{KezerasPRD2020} using A and B sets for the potential parameters listed in Table \ref{tab-1R}. The comparison of the widths obtained in the $ABC$ and
$AAC$ models shows that they are close enough with a negligible difference about 1 - 2 MeV: $\Gamma _{KK\bar{K}}=104-106$ MeV and $\Gamma _{KK\bar{K}}=117-119$ MeV for the potential with A and B parameter sets, respectively.

In this work we study the resonance $K(1460)$ considering only the channel $KK\bar{K}$. One could also have channels like $\pi\pi K$ and $\pi\eta K$. These channels are included, in an effective way, in \cite{Albaladejo}, using Faddeev equations in momentum space by Martinez Torres et al. \cite{Torres2011}, and employing the complex-scaling method \cite{Dote2015} in the semi-relativistic framework in Ref. \cite{SHY19}. While the formation of the resonance $K(1460)$ could be due to the $KK\bar{K}$ channel mainly, the inclusion of other channels may have a relevant role in its mass and, more importantly, in its width. However, consideration of these channels in \cite{Torres2011} gives a significantly smaller width ($\Gamma =50$ MeV) than our single-channel result.  Resonance positions in three-channel $KK\bar K - \pi\pi K - \pi\eta K$ and two-channel $KK\bar K - \pi\pi K$ and $KK\bar K - \pi\eta K$ calculations presented in \cite{SHY19} demonstrate that the coupling to the $\pi\pi K$ channel is significant to reproduce the large width of the resonance and the coupling to $\pi\eta K$ channel makes a large contribution to the mass of the resonance. However, the consideration of  the coupled-channel $\pi\pi K$ and $\pi\eta K$ and variation of the interaction range and strength of the one-range Gaussian potentials did not reproduce the experimental width. In Ref. \cite{Parganlija2017} the channels $K^{*}_{0}(1430)\pi$, $K\rho$, and $K^{*}_{0}(892)\pi$ are quoted as “decaying channels”. These channels require two-body
dynamics either beyond $s-$wave (to form $K^{*}(892)$ or $\rho$) or well above 1 GeV (to form $K^{*}(1430)$), and
these effects are not 
included in \cite{Albaladejo,Torres2011,SHY19}. It is then natural that the width reported in those and
present works is much smaller than the one quoted indicated in  \cite{Brandenburg,DAUM1981} ($\Gamma \sim 250$ MeV) or in the recent LHCb analysis  \cite{LHCb} ($\Gamma \sim 335$ MeV).

\begin{table}[t]
\caption{ The binding energy of the $K^{0}K^{+}K^{-}$ and $K^{+}K^{0}\bar{K}^{0}$ particle configurations in the $AAC$ and $ABC$ model. Calculations are performed for the
potentials with the A and B parameter sets, respectively. In the $AAC$ model, the masses of the $K^{+}$
and $K^{0}$ kaons are equal to the average value of their masses ($\overline{m}_{K}=495.7$ MeV) and the Coulomb attraction is omitted.
A$_{C}$ and B$_{C}$ correspond to the calculations in the $ABC$ model when the Coulomb attraction is
included. The $E_{3}$ is the binding energy of the $KK\bar{K} $ system and $E_{3}(V_{K^{0}K^{+}}=0)$ is the three-body energy, when the repulsive interaction
between $K^{+}$ and $K^{0}$ is omitted. The $E_{2}^{K\bar{K}}$ and $E_{2}^{K^{+}\bar{K}}$, $E_{2}^{K^{0}\bar{K}}$ are the energy of the bound $K\bar{K}$ and  $K\bar{K}$, $K^{0}\bar{K}$ pairs, in $AAC$ and $ABC$ models, respectively.
The energies are given in MeV.}
\centering
\begin{tabular}{@{}cccccccc}
\hline\hline
\multicolumn{8}{c}{$AAC$\ model} \\ \hline
\multicolumn{8}{c}{%
\begin{tabular}{@{}ccccccc}
\noalign{\smallskip} Resonance & System &  \ \ Mass, MeV  & Potentials & $\
\ E_{3}$ \  & $E_{3}(V_{K^{0}K^{+}}=0)$ & $\ E_{2}^{K\bar{K}}$ \\ \hline
${K^{0}}(1460)$ & $K^{0}K^{+}K^{-}$ & \multicolumn{5}{c}{%
\begin{tabular}{ccccc}
$m_{K^{-}}=493.7$ & \ \ \ \ \ A \ \ \ \ \  & $-19.4$ & $\ \ \ \ \ -31.2\ \ \
\ \ $ & $-10.95$ \\
$\overline{m}_{K}=495.7$ & B & $-21.9$ & $-28.9$ & $-11.03$%
\end{tabular}%
} \\ \hline
${K^{+}}(1460)$ & $K^{0}K^{+}\overline{{K}^{0}}$ & \multicolumn{5}{c}{%
\begin{tabular}{ccccc}
$\overline{m}_{K}=495.7$ & \ \ \ \ \ A \ \ \ \  & $-20.1$ & $\ \ \ \ \
-32.1\ \ \ \ \ $ & $-11.40$ \\
$m_{\overline{{K}^{0}}}=497.6$ & \ B & $-22.4$ & $-29.6$ & $-11.34$%
\end{tabular}%
}%
\end{tabular}%
} \\ \hline
\multicolumn{8}{c}{$ABC$ model} \\ \hline
\noalign{\smallskip} Resonance & System & \ \ \ Mass, MeV \ \  & \ \
Potentials \ \  & $\ \ \ E_{3}$ \ \  & \ $\ E_{3}(V_{K^{0}K^{+}}=0)$ & $\ \
\ E_{2}^{K^{+}\bar{K}}$ \ \  & $E_{2}^{K^{0}\bar{K}}$ \\ \hline
${K^{0}}(1460)$ & $K^{0}K^{+}K^{-}$ & \multicolumn{6}{c}{%
\begin{tabular}{c}
$m_{K^{-}}=493.7$ \\
$m_{K^{+}}=493.7$ \\
$m_{K^{0}}=497.6$%
\end{tabular}%
\begin{tabular}{ccccc}
\ \ \ \ \ \ \ \ \ \ A \ \ \ \ \ \ \  & $\ -19.4$ & $\ \ \ \ \ \ \ \ -31.2\ \
\ $ & $-10.74$ \  & $-11.17$ \\
\ \ \ B & $-21.9$ & $\ \ \ \ -28.9$ & $-10.86$ & $-11.17$ \\
\ \ \ \ \ A$_{C}$ & $-20.9$ & $-$ & $-12.37$ & $-11.17$ \\
\ \ \ \ \ B$_{C}$ & $-23.2$ & $-$ & $-12.27$ & $-11.17$%
\end{tabular}%
} \\ \hline
${K^{+}}(1460)$ & $K^{0}K^{+}\overline{{K}^{0}}$ & \multicolumn{6}{c}{%
\begin{tabular}{c}
$\ \ m_{K^{+}}=493.7$ \\
\ $m_{K^{0}}=497.6$ \\
$\ m_{\overline{{K}^{0}}}=497.6$%
\end{tabular}%
\begin{tabular}{ccccc}
\ \ \ \ \ \ \ \ \ A \ \ \ \ \ \ \ \  & $\ -20.1$ & $\ \ \ \ \ \ \ -32.1\ \ \
$ & $-11.17$ & $-11.62$ \\
\ B & $\ -22.4$ & $\ \ -29.6$ & $-11.17$ & $-11.49$%
\end{tabular}%
} \\ \hline\hline
\end{tabular}
\label{tab-3R}
\end{table}

\begin{figure}[t]
\centering
\includegraphics[width=18.5pc]{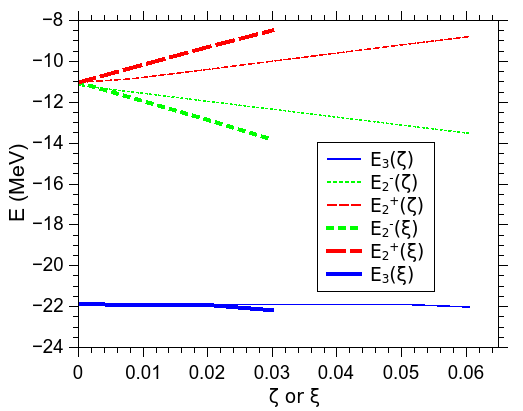} %
\caption{ The bound state energy $E_{3}(\protect\zeta )$
of the $K^{0}K^{+}K^{-}$ system (thin line) and the binding energies of
bound $\ E_{2}^{K^{0}K^{-}}$ ($E_{2}^{+}(\protect\zeta )$) and $%
E_{2}^{K^{+}K^{-}}$($E_{2}^{-}(\protect\zeta )$) kaon pairs (thin dashed
lines) as functions of the mass scaling parameter $\protect\zeta $. The
dependencies of bound state energy $E_{3}(\protect\xi )$ (thick line) and the
binding energies of bound $\ E_{2}^{K^{0}K^{-}}$ ($E_{2}^{+}(\protect\xi )$)
and $E_{2}^{K^{+}K^{-}}$($E_{2}^{-}(\protect\xi )$) kaon pairs (thick dashed
lines) in the $K^{0}K^{+}K^{-}$ system on the potential scaling parameter $%
\protect\xi $. Calculations are performed with the set B for $KK$
and $K{\bar{K}}$ interactions.}
\label{fig6b}
\end{figure}

\begin{figure}[t]
\centering
\includegraphics[width=17.pc]{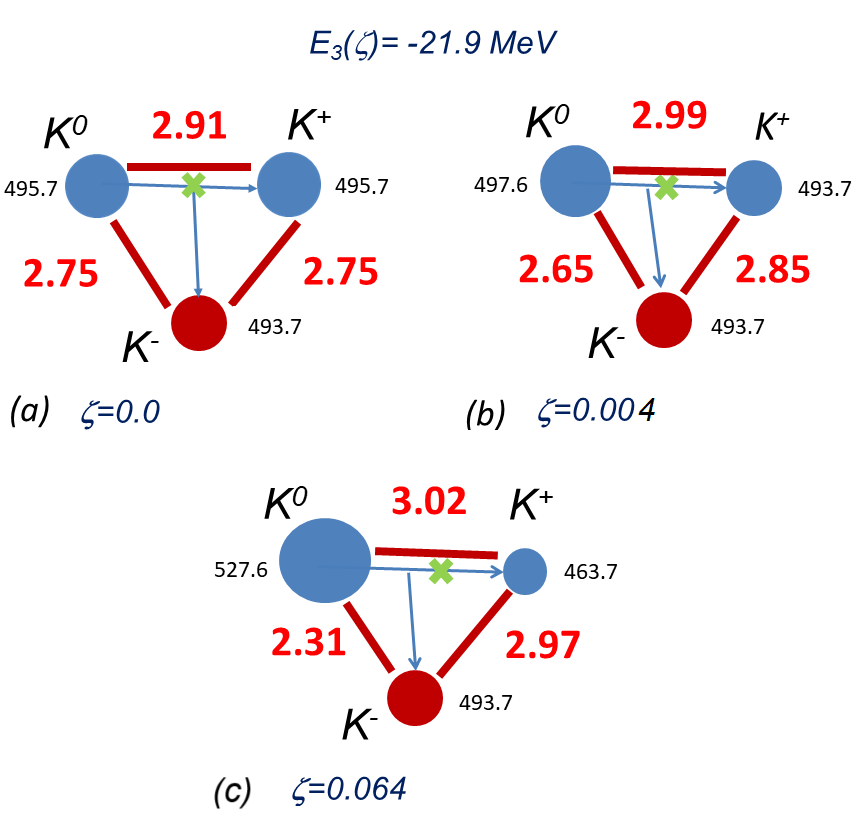}
\includegraphics[width=18.5pc]{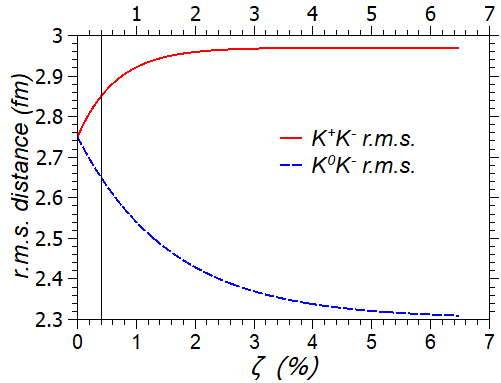}
\caption{(\textbf{Left panel}) The mass
redistribution effect on the root mean squared distances between kaons in the $%
K^{0}K^{+}K^{-}$ system. The results of calculations for the  $AAC$ model ($a$), the $ABC$
model using the experimental kaon masses ($b$), the $ABC$ model with the
different unrealistic masses for the $K^{0}$ and $K^{+}$ kaons ($c$). The
total mass of the $K^{0}K^{+}$ pair is constant and equals to the sum of experimental
masses. In ($a$)-($c$) cases the total binding energy $%
E_{3}(\protect\zeta )=-21.9$ MeV is the same, while the $r.m.s.$
distances between kaons are different.
(\textbf{Right panel}) The $r.m.s.$ distances between $K^{+}K^-$ (solid curve) and $K^{0}K^-$ (dashed curve) kaons in
three-body systems as functions of the $\zeta$ parameter. The vertical line corresponds to the parameter $\zeta$ related to experimental masses of $K^{+}$ and $K^{0}$ kaons.
Calculations are performed with the set B for $KK$
and $K{\bar{K}}$ interactions.
}
\label{fig6c}
\end{figure}

\begin{figure}[h]
\centering
\includegraphics[width=16.pc]{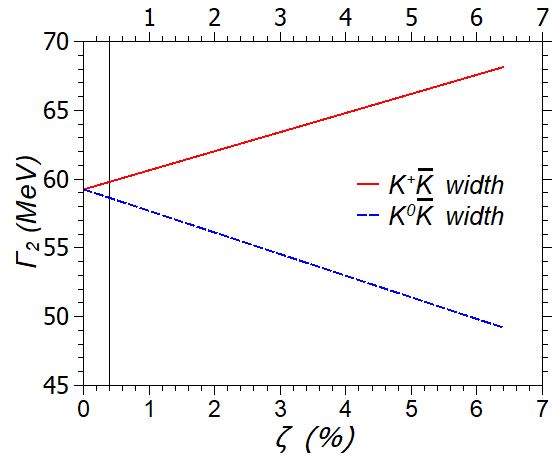}
\caption{
The two-body widths $\Gamma_2$ of $K^{+}K^-$ (solid line) and $K^{0}K^-$ (dashed line) versus the $\zeta$ parameter. Calculations are performed with the set B for $K{\bar{K}}$ interaction. The vertical line shows the parameter $\zeta$ related to experimental values for masses of $K^{+}$ and $K^{0}$ kaons.
}
\label{fig6d}
\end{figure}

\subsection{The mass-symmetry breaking in the $AA$ pair: 
from the $AAC$ to $ABC$ model}

Let us start with the bosonic $AAC$ model with two identical particles with
the average mass of two kaons and violate the mass-symmetry of this model by
changing the masses of two identical particles but keeping the total mass of
the $AA$ pair constant. Such mass redistribution leads to the transformation
of the $AAC$ model with the symmetric wave function with respect to the
exchange of $AA$ particles to the $ABC$\ model with a lack of this symmetry.
Now, in the $AA$ pair of  ''identical'' particles, we have the masses $$m_{K^{0}}(\zeta)=\overline{m}%
_{K}(1-\zeta ), \qquad m_{K^{+}}(\zeta)=\overline{m}_{K}(1+\zeta ),$$ where $%
\overline{m}_{K}=\left( m_{K^{0}}+m_{K^{+}}\right) /2$ the average mass of
the $AB$ pair and $\zeta $ is a mass scaling parameter that can be varied.
The total mass of this pair is constant: $%
m_{K^{0}}(\zeta)+m_{K^{+}}(\zeta)=m_{K^{0}}+m_{K^{+}}$. The kaonic system with the mass $m_{K^{-}}$ and variable masses $%
m(K^{0})$ and $m(K^{+})$, must be considered within the $ABC$
model. 
The cases when $%
\zeta =0$ and $\zeta =0.004$ correspond to the $AAC$ model with average
masses of $K^{0}$ and $K^{+}$ kaons and the $ABC$ model that describes $%
K^{0}K^{+}K^{-}$ with the experimental masses of kaons, respectively.

Results of calculations of the binding energy within the $ABC$\ model with
variable masses of two kaons and the energy of the bound $AC$ and $BC$ kaon
pairs as functions of the mass scaling parameter $\zeta $\ are shown in
Fig. \ref{fig6b}. 
The total energy $E_{3}(\zeta )$ does not
depend on the mass redistribution between $A$ and $B$ kaons up to $\zeta \le$ 0.06.
The later value shows when the limit of  the approximation (\ref{D1}) is reached.
 However, the
bound $AC$ and $BC$ kaon pairs energies are sensitive to the variation of
the parameter $\zeta $. The $K^{+}K^{-}$ and $K^{0}K^{-}$ kaon pairs energies $E_{2}^{+}(\zeta )$ and $E_{2}^{-}(\zeta )$, which correspond to the
increase of the $K^{0}$ mass and the decrease
of the $K^{+}$ mass from the average value to the experimental
mass, increases and decreases, respectively, with the $\zeta $ increase. 
Thus, the redistribution of the mass
between two kaons violates the exchange symmetry of the wave function in the
$AAC$ model and leads to the $ABC$ model that gives the same total energy as
the $AAC$ model, but increases $E_{2}^{+}(\zeta )$ and decreases $%
E_{2}^{-}(\zeta )$ energies of the bound kaon pairs. The violation of the
wave function symmetry from the symmetric with respect to the exchange of $AA
$ particles to a wave function without such symmetry should affect the
average distance between kaons.
To demonstrate that we calculated the root mean
squared ($r.m.s.$) distances between kaons.  
The latter is illustrated in Fig. \ref{fig6c}, the left panel, by presenting the $r.m.s.$ distances between kaon pairs for the $%
K^{0}K^{+}K^{-}$ particle configuration for the different values of $\zeta$. In the $AAC$ model $\zeta = 0$ and one gets the isosceles triangle. Consideration of experimental masses in
the $ABC$ model leads to the different root mean squared distances between particles. In Fig. \ref{fig6c}, the right
panel, is shown the dependence of the $r.m.s.$ distances on $\zeta$.
One can conclude that the mass flow affects the $r.m.s.$   distances between kaons up to $\zeta \sim 0.06$. This effect is non-linear and tends to saturate. 

According to Eq. (\ref{Jc}) and Eq. (\ref{e:1}), the mass-scaled
coordinates in the $AAC$ model are $\mathbf{x}_{1}=\sqrt{\overline{m}_{K}}%
\mathbf{r}_{23}$ and $\mathbf{x}_{2}=\sqrt{\frac{2\overline{m}_{K}m_{\bar{K}}%
}{\overline{m}_{K}+m_{\bar{K}}}}\mathbf{r}_{13}$. These coordinates in the $%
ABC$ model are $\mathbf{x}_{1}=\sqrt{\frac{\overline{m}_{K}^{2}-(\Delta
m)^{2}}{\overline{m}_{K}}}\mathbf{r}_{23}$, $\mathbf{x}_{2}=\sqrt{\frac{2(%
\overline{m}_{K}+\Delta m)m_{\bar{K}}}{\overline{m}_{K}+\Delta m+m_{\bar{K}}}%
}\mathbf{r}_{13}$, $\mathbf{x}_{3}=\sqrt{\frac{2(\overline{m}_{K}-\Delta
m)m_{\bar{K}}}{\overline{m}_{K}-\Delta m+m_{\bar{K}}}}\mathbf{r}_{12}$ 
and have a $\Delta m$%
-dependence. Thus, the dependence of pair potentials as a function of the $%
\mathbf{x}$ coordinate in Eq. (\ref{e:1}) is expressed as a $\Delta m$%
-dependence.

One would assume that three-body kinetic energy operator is affected by the mass scaling parameter $\zeta$. However, in the first
order perturbation theory for $\zeta \le 0.064$, as we have shown in Section IV, the kinetic energy matrix element $\langle {\hat{H_{0}}}\rangle$ does not depend on $\zeta$.
The binding energy $E=\langle {\hat{H_{0}}}\rangle+\langle (v_{K^{0}\bar{K}}+v_{K^{+}\bar{K}}+v_{KK})\rangle$ is also a constant. This means the matrix element of the total potential energy does not changed with the $\zeta$ increasing.
Because  the width $\Gamma$ is evaluated from the imaginary part of the complex potentials,
the three-body width is also independent on the parameter $\zeta$ and is the same in the $AAC$ and $ABC$ models. 
The two-body $K^{+}\bar K$ and $K^{0}\bar K$ subsystems do not have the compensation mechanism due to the absence of the third particle that has the three-body $KK\bar K$ system as expressed by Eqs. (\ref{D})-(\ref{D1}).
Two-body widths depend on the mass scaling parameter and repeat the $E_2(\zeta)$ dependence as depicted in Fig.  \ref{fig6d}. The widths of the $K^{+}\bar K$ and $K^{0}\bar K$ pairs are the same, $\Gamma_{2}=59.2$ MeV, in the $AAC$ model ($\zeta=0$). In the $ABC$ model, $\zeta=0.004$, the widths are $\Gamma_{2}=59.8$ MeV and $\Gamma_{2}=58.6$ MeV for the $K^{+}\bar K$ and $K^{0}\bar K$, respectively, for the set of the parameters B.

\subsection{From the $AAC$ to $ABC$ model through asymmetry of $AC$ and $BC$
potentials}

Above we considered the bosonic $AAC$ model and its transformation to the $%
ABC$ model due to changing the masses of two identical particles. However,
the bosonic $AAC$ model can be also transformed to the $ABC$ model in the case when
interactions in the $AC$ and $BC$ pairs are different. For example, in Ref. \cite{SHY19} to interpret the three-body resonance as K(1460) were used two-body potentials with the different strength and range parameters.
In our consideration of the $%
K^{0}K^{+}K^{-}$ particle configuration the kaon-antikaon interaction we use the same strong interaction in the
$K^{0}K^{-}$ and $K^{+}K^{-}$ pairs. However, interactions in the
$K^{0}K^{-}$ and $K^{+}K^{-}$ pairs are different due to the Coulomb
attraction between  $K^{+}$ and $K^{-}$. The latter means that even if one
considers the average mass for $K^{0}$ and $K^{+}$ kaons, the $K^{0}K^{+}K^{-}$ configuration
should be described within the $ABC$ model. Let us demonstrate this in general via a model where a strong interaction in the $AC$ and $BC$ pairs is different
by introducing scaled potentials
$$
v_{AC}\to v_{K^{0}K^{-}}(\xi)=\overline{v}%
(1-\xi ),\qquad v_{BC}\to v_{K^{+}K^{-}}(\xi)=\overline{v}(1+\xi ).
$$
 In the last expressions $\overline{v}$ is
the average potential of the $AC$ and $BC$ pairs, $\xi $ is the potential scaling parameter and $v_{AC}=v_{BC}$ when $\xi = 0$. Within this model we
calculate the binding energies of the $K^{0}K^{+}K^{-}$ and the bound pairs $%
K^{0}K^{-}$ and $K^{+}K^{-}$ by keeping the average potential constant.
In Fig. \ref{fig6b} are shown dependencies of the binding energy $E_{3}(\xi )
$ of the $K^{0}K^{+}K^{-}$ and energies $\ E_{2}^{K^{0}K^{-}}$ ($%
E_{2}^{+}(\xi )$) and $E_{2}^{K^{+}K^{-}}$($E_{2}^{-}(\xi )$) of the $%
K^{0}K^{-}$ and $K^{+}K^{-}$ pairs, respectively, on the potential scaling
factor $\xi $. Results when $\xi =0$ correspond to the $AAC$ model, while
the increment of $\xi $ leads to the $ABC$ model. The increase of $\xi $
leads to the increase or decrease of the two-body energies of the $K^{0}K^{-}
$ and $K^{+}K^{-}$ pair, respectively. At the same time, the binding energy
of the $K^{0}K^{+}K^{-}$ system remains unchanged up to a value of $\xi =0.03$,
as is demonstrated in Fig. \ref{fig6b} and can be seen from Eq.
(\ref{eq:22a}). A similar situation was demonstrated in the previous
subsection when the mass scaling parameter $\zeta $\ increases up to the
value of $\zeta =0.06$.

The Coulomb attraction acting in the single kaon-antikaon pair leads also to
the $ABC$ model and increases three-body energy by 1.5 MeV and 1.3 MeV for
the parameter sets A and B for $KK$ and $K{\bar{K}}$ interactions,
respectively. It can be noted that the averaging procedures presented in
sections \ref{sec:3} and \ref{sec:4} allow us to consider the Coulomb
interaction in the $AAC$ model.

\section{Conclusions}

\label{sec:6} In the framework of the Faddeev equations in configuration
space, we investigated the $K$(1460) resonance dynamically generated via the
$KK\bar{K}$ system. We considered the $KK{\bar{K}}$ system as the $%
K^{0}K^{+}K^{-}$ and $K^{0}K^{+}\overline{{K}^{0}}$ particle configurations
that are analyzed using $AAC$ and $ABC$ models. We demonstrated that the $ABC
$ model can be reduced to the $AAC$ one, where the wave function is
symmetric with respect to the exchange of identical particles. The reduction
is possible by averaging the masses of the $AB$ pair or averaging $AC$ and $%
BC$ potentials, if they are different.

It is shown that the repulsive $KK$ interaction plays essential role in the binding energy of the $KK\bar K$ system: contribution of the repulsive $KK$
interaction decreases the three-particle binding energy by about 38\% and 25\% for the A  and B parameter sets, respectively.

Our three-body non-relativistic single-channel model predicts a quasi-bound
state for the $KK\bar{K}$ system. The mass of neutral $K(1460)$ resonance
calculated in the $ABC$\ model for the $K^{0}K^{+}K^{-}$ particle
configuration is 1464.1 MeV or 1461.8 MeV, while the mass of the charged $%
K^{+}(1460)$\ resonance for the ${K}^{0}K^{+}\overline{K^{0}}$ particle
configuration is 1468.8 MeV or 1466.5 MeV. These values are obtained for
the parameter sets A and B for $KK$ and $K{\bar{K}}$ interactions,
respectively. The results are in fair agreement with the
experimental value of the $K$(1460) mass, 1482.40$\pm $3.58$\pm $15.22 MeV
\cite{PDG2022}.

Due to the Coulomb attraction of three-body binding energy $E_{3}$\
increases, and the energy shift is 1.5 and 1.3 MeV for the parameter sets A
and B for $KK$ and $K{\bar{K}}$ interactions, respectively. Let us note that
within the $AAC$\ model, consideration of the Coulomb attraction is also
possible using the averaging procedure for the Coulomb potential. The
binding energy $E_{3}$ of $K^{0}K^{+}K^{-}$ and ${K}^{0}K^{+}\overline{K^{0}}
$ calculated in the $AAC$ model with the average kaon masses and in the $ABC$
model with the experimental kaon masses are the same. Effectively, the $AAC$
model can reproduce the binding energy obtained in the $ABC$ model if the
corresponding relative correction of masses is not larger than 6\%. Thus,
the small difference of the kaon masses does not affect the binding energy
of the $K^{0}K^{+}K^{-}$ kaonic system when the Coulomb interaction is
neglected.

An increase in the mass difference of the kaons leads to the mass-symmetry
violation of the system with the same binding energy. However, when the
relative mass correction exceeds 6\% the binding energies calculated using
the $AAC$ and $ABC$ models differ. It should be noted that this effect has
not been reported before. The three-body kaonic system allows us to
demonstrate this effect clearly. One can consider similar nuclear systems,
for example, $np\Lambda $ or $np\alpha $, instead of the $KK\bar{K}$ kaonic
system but the effect will be small due to a small difference between the
proton and neutron masses. We have found that mass correction does not
change three-body energy, however, violates the exchange symmetry that
affects the $ABC$ model wave function symmetry.

In the $AAC$ model, there is the exchange symmetry, related to the
symmetric localization of identical particles. In contrast to the $AAC$
model, consideration of experimental masses in the $ABC$ model leads to the
violation of this symmetry. This indicated by different root mean squared
distances between kaons. The different $r.m.s.$
distances between kaons are due to different kaon masses and/or potentials
in the $AAC$ and $ABC$ models.

We considered the $KK\bar{K}$ system using the single-channel description
with effective $s$-wave potentials. Some refinements can be done, such as
using more realistic two-body potentials, including $p$-wave components,
and/or considering the coupled-channel approach. 
It is important to note that the choice of the model and
assumptions made in our analysis can always have an impact on the results.
Therefore, it is essential to carefully consider the limitations and
uncertainties for description $KK\bar{K}$ system relating to the mass and charge symmetry braking. 

\section*{Acknowledgments}

This work is supported by the National Science Foundation grant HRD-1345219
and DMR-1523617 awards Department of Energy$/$National Nuclear Security
Administration under Award Number NA0003979 DOD-ARO grant \#W911NF-13-0165.

\end{document}